\documentclass[%
 reprint,
superscriptaddress,
 amsmath,amssymb,prl
 aps,
floatfix,
]{revtex4-2}

\usepackage{graphicx}
\usepackage{dcolumn}
\usepackage[colorlinks=true,allcolors=blue]{hyperref}
\usepackage{bm}

\begin{document}

\preprint{APS}

\title{Thermodynamic inference of correlations in nonequilibrium collective dynamics}
\author{Michalis Chatzittofi}
\address{Department of Living Matter Physics, Max Planck Institute for Dynamics and Self-Organization, D-37077 Göttingen, Germany}
\author{Ramin Golestanian}
\email{ramin.golestanian@ds.mpg.de}
\address{Department of Living Matter Physics, Max Planck Institute for Dynamics and Self-Organization, D-37077 Göttingen, Germany}
\address{Rudolf Peierls Centre for Theoretical Physics, University of Oxford, Oxford OX1 3PU, United Kingdom}
\author{Jaime Agudo-Canalejo}
\email{j.agudo-canalejo@ucl.ac.uk}
\address{Department of Living Matter Physics, Max Planck Institute for Dynamics and Self-Organization, D-37077 Göttingen, Germany}
\address{Department of Physics and Astronomy, University College London, London WC1E 6BT, United Kingdom}
\date{\today}

\begin{abstract}
The theory of stochastic thermodynamics has revealed many useful fluctuation relations, with the thermodynamic uncertainty relation (TUR) being a theorem of major interest. When many nonequilibrium currents interact with each other, a naive application of the TUR to an individual current can result in an apparent violation of the TUR bound. Here, we explore how such an apparent violation can be used to put a lower bound on the strength of correlations {$C$} as well as the number {$N$} of interacting currents  in collective dynamics. {This lower bound is a combined bound on $C(N-1)$ if only one current is measured, or a bound on $N$ if two currents are measured.} Our proposed protocol allows for the inference of hidden correlations in experiment, for example when a team of molecular motors pulls on the same cargo but only one or a subset of them is fluorescently tagged. By solving analytically and numerically several models of many-body nonequilibrium dynamics, we ascertain under which conditions this strategy can be applied and the inferred bound on correlations becomes tight.
\end{abstract}

\maketitle

Entropy production rate (EPR) is the measure of nonequilibrium activity in a stochastic system  and is tied to the existence of nonequilibrium currents in the system \cite{seifert2019stochastic,nardini2017entropy}. The thermodynamic uncertainty relation (TUR) \cite{barato} quantifies the trade-off between EPR and the precision of the nonequilibrium currents, where precision is related to the ratio between the average and the standard deviation of the fluctuating currents. The TUR, which has been proven rigorously \cite{gingrich2016dissipation,godec} and has been confirmed in experiments \cite{sperms,pal2020experimental}, has found its most practical application in the inference of (lower bounds for) nonequilibrium driving forces given experimental measurements of fluctuating currents \cite{inferring2020,Skinner2021}. Relevant experimental systems include active matter \cite{marchetti}, molecular machines \cite{Chatzittofi2024EPL} such as motors \cite{RevModPhys.69.1269} and enzymes \cite{agudo2021synrhonization}, stochastic oscillators \cite{kuramoto1984chemical,Chatzittofi2023,lee2018thermodynamic,barato2016cost}, microscopic heat engines \cite{PhysRevLett.120.190602}, artificial  nanorotors  \cite{pumm2022dna,Shi2023}, and even in open quantum systems  \cite{PhysRevLett.126.010602,PhysRevLett.126.210603}.  Additionally, the TUR has also inspired other important thermodynamic relations  \cite{macieszczak,horowitz2020thermodynamic,PhysRevE.108.054126}, placing bounds on e.g.~the extent of anomalous diffusion  \cite{anomalous2021hartich}, the asymmetry of cross-correlations \cite{asymmetry_cross}, and correlation times  \cite{correlation_times}.

In its original form, the TUR for a nonequilibrium system in steady state was proposed for a scalar (one dimensional) fluctuating current and can be expressed as \cite{barato}
\begin{align} \label{eq:tur}
    \mathcal{J}^2/D_\mathcal{J}\leq \dot \sigma /k_B,
\end{align}
where $\mathcal{J}$ represents the steady state average of the scalar observable current of interest, $D_\mathcal{J}$ the diffusion coefficient associated to the corresponding fluctuating observable, $\dot \sigma$ the steady state average EPR, and $k_B$ the Boltzmann constant. An important generalization to vectorial fluctuating currents, or equivalently to several scalar currents that are simultaneously observed, is the multidimensional thermodynamic uncertainty relation (MTUR) given by \cite{dechant}
\begin{align}\label{eq:mtur}
\bm{\mathcal{J}}^T \cdot \bm{\mathcal{D}}^{-1} \cdot \bm{\mathcal{J}} \leq  \dot \sigma/k_B,
\end{align}
where $\bm{\mathcal{J}}$ is now the steady state average of the vectorial observable current and $\bm{\mathcal{D}}$ is the covariance tensor associated with the fluctuating multidimensional observable. The MTUR allows for the inference of tighter lower bounds on the entropy production in systems with multiple degrees of freedom, when more than one observable can be tracked simultaneously, for example in interacting many-body systems  \cite{manybody}. Alternatively, one may use the MTUR together with known mechanistic information about the coupling between degrees of freedom to obtain tighter bounds on dissipation even when only one observable is tracked, as recently proposed for stochastic swimmers with coupled chemical and mechanical degrees of freedom \cite{chatzittofi2023entropy}.

 \begin{figure}[ht]
     \centering
     \includegraphics[scale=0.055]{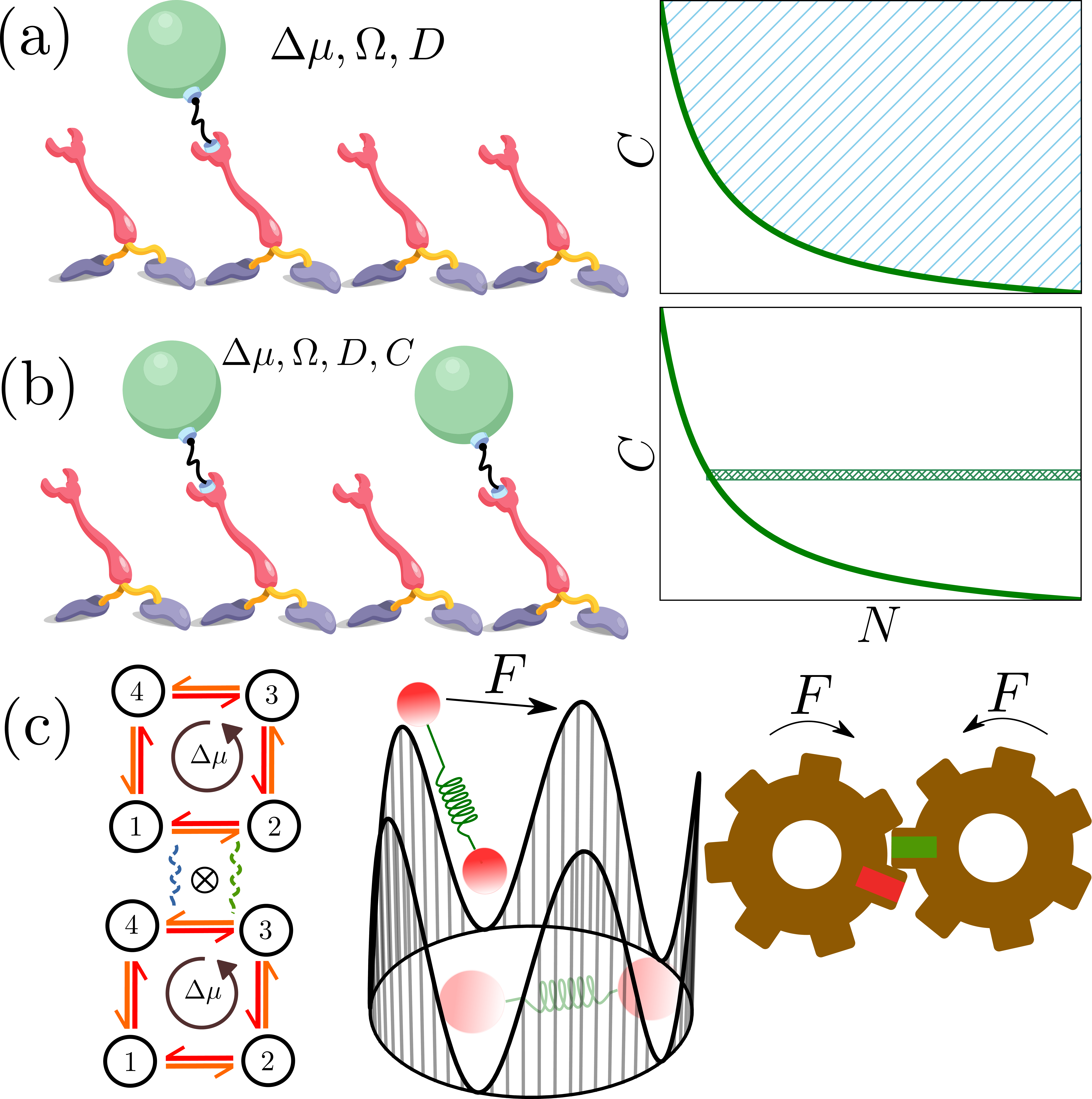}
     \caption{(a) Measurement of a single observable (here, the position of a fluorescently-tagged molecular motor) allows for inference of a combined bound on the strength of correlations $C$ and the number of interacting processes $N$. (b) Simultaneous measurement of two observables fixes the strength of correlations and places a bound on the number of interacting processes $N$. (c) Examples of coupled identical processes: two discrete biomolecular processes driven by a cyclic affinity $\Delta \mu$;  two colloids in an optical ring driven by a constant force $F$; two molecular gears driven by a constant force $F$.}
     \label{fig:figure_one}
\end{figure}

In this Letter, we propose to turn the MTUR on its head and exploit it to infer the existence of hidden correlations in a system, even when only a single observable is experimentally accessible. We show that this is possible in systems satisfying two simple conditions: (i) many statistically-identical processes interact with each other, and (ii) the observable quantities are tightly coupled to entropy production, with a known rate of entropy production per step. This may for example represent ensembles of identical molecular motors walking on the same biofilament \cite{klumpp2005cooperative,costantini2024thermodynamic}, clustered enzymes catalyzing chemical reactions in a metabolon \cite{sweetlove2018role,o2021role}, clustered rotors or channels in a membrane \cite{jimenez2014mitochondrial,visscher2017supramolecular}, or driven colloids in an optical ring \cite{lee2006giant,hydrodynamic2011sokolov,PhysRevE.90.042302}. We will first introduce the general strategy, valid for any system that satisfies the two conditions just described. We will then study two toy models that are analytically solvable, and two models that we solve numerically, in order to ascertain under which conditions the proposed strategy can be applied, and when does the inferred bound on correlations become tight.

\textit{Inference of correlations.---}
We consider $N$ stochastic processes that are identical, in the sense that they are governed by the identical underlying stochastic dynamics, and are all-to-all coupled in a statistical sense, i.e.~will show identical pair correlations with each other after a sufficiently long observation time (as expected in an ergodic system). Let us denote the associated scalar observables as $(\phi_1,..., \phi_N)$. Quantitatively, the conditions just described imply that all scalar observables have the same average current $\Omega \equiv \lim_{t \to \infty} \langle  \phi_i \rangle/t$, the same diffusion coefficient $D\equiv \lim_{t \to \infty} (\langle \phi_i^2 \rangle-\langle \phi_i \rangle^2)/(2t)$, and the same pair correlation strength $C \equiv \lim_{t \to \infty}  (\langle \phi_i \phi_j \rangle - \langle \phi_i\rangle \langle \phi_j \rangle)/\sqrt{(\langle \phi_i^2 \rangle-\langle \phi_i \rangle^2)(\langle \phi_j^2 \rangle-\langle \phi_j \rangle^2)}$ (for $i\neq j$). Note that $C$ is bounded between $-1/(N-1)$ for maximally anticorrelated processes and $+1$ for perfectly correlated processes.
Lastly, we assume that the observable currents are driven by energy dissipation (entropy production) through a tightly coupled mechanism \cite{barato,pietzonka2016universal,barato2016cost,lee2018thermodynamic,marsland2019thermodynamic,manybody,leighton2022dynamic} so that, for every individual current, we can write an average energy dissipation rate that is proportional to the average current $\dot{\sigma}^{(1)} T \equiv \Omega \Delta \mu$, with $\Delta \mu$ the energy dissipated per step {and $T$ the temperature of the bath}. The total EPR in the system is then $\dot{\sigma} = N \dot{\sigma}^{(1)}$.

With these choices, application of the MTUR (Eq.~\eqref{eq:mtur}) and a rearrangement of the terms result in the inequality
\begin{equation}\label{eq:ctur}
\frac{\Omega}{D}\frac{k_B T}{\Delta \mu} - 1 \leq C(N-1),
\end{equation}
which puts a lower bound on the correlation strength $C$ (and the number  of interacting processes $N$) given a measurement of the average current $\Omega$ and the diffusion coefficient $D$, and provided that the dissipation per step $\Delta \mu$ is known. We note that Eq.~\eqref{eq:ctur} can also be obtained by applying the standard TUR (Eq.~\eqref{eq:tur}) to the observable corresponding to the total sum $\sum_i \phi_i$.

To get an intuition for the meaning of Eq.~\eqref{eq:ctur}, it is useful to note that its left hand side represents a measure of the violation of a naively applied single-current TUR. Indeed, for a single isolated or non-interacting current, the standard TUR [Eq.~\eqref{eq:tur}] gives $\frac{\Omega}{D}\frac{k_B T}{\Delta \mu}  \leq  1$ (consistent with Eq.~\eqref{eq:ctur} with $N=1$ or $C=0$). Thus, if measurement of a single observable appears to violate (outperform) this naive TUR, it implies that the left hand side of Eq.~\eqref{eq:ctur} is positive, and therefore that there must be positive correlations in the system ($C>0$ and $N \geq 2$). If, on the other hand, the naive TUR is satisfied, it means that the measurement is compatible with the absence of correlations in the system, and Eq.~\eqref{eq:ctur} only serves to rule out negative correlations stronger than those allowed by the bound.

When a single observable is tracked, e.g.~when only one molecular motor within a team is fluorescently labeled, Eq.~\eqref{eq:ctur} puts a combined lower bound on the correlation strength $C$ and number of interacting processes $N$, see Fig.~\ref{fig:figure_one}(a). If  one can additionally measure the correlation strength, e.g.~if two or more motors within the team are labeled, one can infer a lower bound on the number of processes $N$, see Fig.~\ref{fig:figure_one}(b). In the following, we present several minimal models [Fig.~\ref{fig:figure_one}(c)] that allow us to ascertain the conditions under which the naive TUR is broken and the proposed strategy can be applied, and those for which the inferred bound of Eq.~\eqref{eq:ctur} becomes tight.

 \begin{figure*}
     \centering
     \includegraphics[scale=0.62]{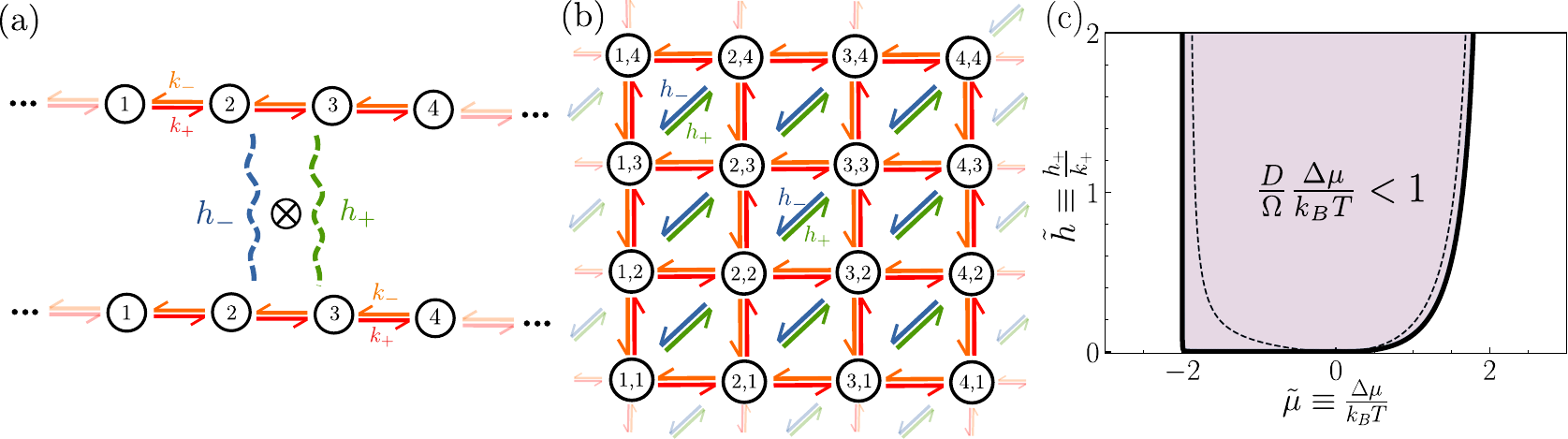}
     \caption{(a) Two identical discrete processes modelled as biased one-dimensional random walks, with red (orange) arrows indicating forward (backward) transitions with rates $k_\pm$. Interactions between the two processes are represented by the green and blue squiggles. (b) The ``outer product'' of the two one-dimensional processes corresponds to a two-dimensional lattice, where the tuples indicate the internal state of the whole system. Interactions are governed by the green and blue arrows, representing simultaneous forward and backwards transitions for both processes with rates $h_\pm$. (c) Parameter space spanned by the nonequilibrium driving force $\tilde{\mu} \equiv \Delta\mu/k_B T$ and the coupling strength $\tilde{h} \equiv h_+/k_+$, showing the regime in which the naive TUR is broken, and thus the existence of nonzero correlations could be inferred in experiment, for a large number of interacting processes ($N \gg 1$). The corresponding boundary for $N=2$ is shown as the dashed line.}
     \label{fig:discrete}
\end{figure*}

\textit{Discrete coupled processes.---} We first consider a rather generic example of coupled discrete Markov processes, which might represent various interacting or coupled biomolecular processes such as molecular motors walking on a track, or chemical reactions catalyzed by nearby enzymes or different monomers in a multimeric enzyme.
As an example, in Fig.~\ref{fig:discrete}(a) we show two identical one-dimensional processes where the red (orange) arrow indicates the forward (backward) rate $k_+$ ($k_-$). For local detailed balance to be satisfied one must impose that $k_+/k_- = e^{\Delta \mu / k_B T}$, where $\Delta \mu$ represents the energy dissipated per transition. The simultaneous dynamics of the two processes can alternatively be viewed as taking place on a two-dimensional lattice of Markov states as shown in Fig.~\ref{fig:discrete}(b), where transitions of one process or the other correspond to hopping horizontally or vertically on the two-dimensional lattice.  To include interactions between the two processes, coupling rates $h_\pm$ are introduced which are represented by the green and blue squiggles in Fig.~\ref{fig:discrete}(a) and arrows in Fig.~\ref{fig:discrete}(b). These correspond to diagonal jumps in the lattice, which imply a forward or backward transition taking place simultaneously for both processes. Since two steps are performed during a coupled transition, detailed balance demands $h_+/h_- = e^{2\Delta \mu / k_B T}$.

The dynamics just illustrated for two coupled processes are straightforwardly extended to $N$ coupled processes, where we assume an all-to-all coupling, such that with rate $h_+$ ($h_-$) all $N$ processes undergo a simultaneous forward (backward) step. In this case, detailed balance demands $h_+/h_- = e^{N\Delta \mu / k_B T}$. Following an analytical derivation (Appendix A), we find that the correlation strength is
 \begin{align}\label{eq:correlation}
     C = \frac{\tilde{h}(1+e^{-N\tilde{\mu}})}{1+e^{-\tilde{\mu}}  + \tilde{h}(1+e^{-N\tilde{\mu}})},
 \end{align}
where $\tilde{h} \equiv h_+/k_+$ is the dimensionless coupling strength and $\tilde{\mu} \equiv \Delta\mu/k_B T$. As may be expected, we find that $C \to 1$ as $\tilde{h} \rightarrow \infty$ and $C = 0$ when {$\tilde h=0$}.
In turn, the ratio of average current to diffusion coefficient can be written as
 \begin{align}\label{eq:dphiom}
    \frac{\Omega}{D}  = 2 \,\frac{1-e^{-\tilde{\mu}}  + \tilde{h}(1-e^{-N\tilde{\mu}})}{1+e^{-\tilde{\mu}}  + \tilde{h}(1+e^{-N\tilde{\mu}})}.
 \end{align}
Combining both expressions, we obtain an exact relation between $\Omega/D$, $C$, and the energy dissipation per step $\tilde{\mu}$, with the form
\begin{align}\label{eq:dphiom2}
     \frac{\Omega}{D} =2 \, \frac{1-e^{-\tilde{\mu}} + 2C\frac{e^{-\tilde{\mu}}-e^{-N\tilde{\mu}}}{1+e^{-N\tilde{\mu}}}}{1+e^{-\tilde{\mu}}}.
\end{align}
This expression can be shown to always satisfy the bound in Eq.~\eqref{eq:ctur}, which it saturates in the near-equilibrium limit $\Delta \mu \rightarrow 0$. In the case where $C=0$ (or $N=1$), the right hand side becomes $2 \tanh(\tilde{\mu} / 2)$, and we recover the relation for the single biased random walk which was used to conjecture the original TUR  \cite{barato}. Our model thus represents the minimal extension of this basic toy model to the case of many interacting processes.

Using Eq.~\eqref{eq:dphiom}, we can investigate under which conditions the naive TUR is violated and $\frac{D}{\Omega}\frac{\Delta \mu}{k_B T} < 1$. For such parameter values, the inference strategies proposed in Fig.~\ref{fig:figure_one}(a,b) can be used to infer the existence of nonzero correlations in the system  and put a lower bound on them. We find that this is possible when the coupling $\tilde{h}$ is larger than a critical coupling strength $\tilde{h}(\tilde{\mu})$, see Fig.~\ref{fig:discrete}(c). Interestingly, this is only possible if the driving forces are weak, with $|\tilde{\mu}| < \tilde{\mu}^*$ where $\tilde{\mu}^* \simeq 1.915$ for $N=2$ and $\tilde{\mu}^* \to 2$ as $N\to \infty$. Indeed, the critical coupling strength diverges as $\tilde{\mu}$ approaches $\pm \tilde{\mu}^*$.

\begin{figure*}
     \centering
    \includegraphics[scale=0.39]{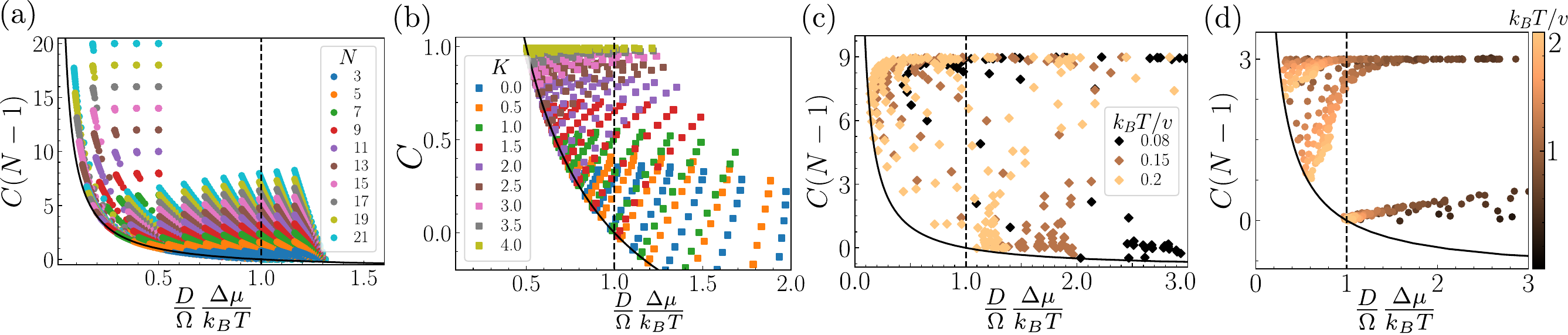}
     \caption{Scatter plots of $\frac{D}{\Omega}\frac{\Delta \mu}{k_B T}$ against $C(N-1)$ for the four different models described in the main text, showing how the bound in Eq.~\eqref{eq:ctur} is satisfied and can be close to saturated. Points to the left of the vertical black dashed line correspond to parameter choices for which $\frac{D}{\Omega}\frac{\Delta \mu}{k_B T} < 1$ and thus the naive TUR is saturated, allowing for the thermodynamic inference of correlations. (a) Discrete model of Fig.~\ref{fig:discrete}; (b) analytically-solvable continuous model; (c) thermally-activated oscillators with dissipative coupling; (d) thermally-activated oscillators with Kuramoto coupling. Parameter choices for each of the four models are described in the text. Simulations in (c) and (d) were performed using Euler-Maruyama integration.}
     \label{fig:figure3}
\end{figure*}

Even further, using Eqs.~\eqref{eq:correlation} and \eqref{eq:dphiom} we can characterize how close to saturation the bound in Eq.~\eqref{eq:ctur} can get, as shown in Fig.~\ref{fig:figure3}(a) where we plot  $\frac{D}{\Omega}\frac{\Delta \mu}{k_B T}$ against $C(N-1)$ for a range of parameter values in $0 \leq \tilde{h} \leq 1$, $-1 \leq \tilde{\mu} \leq 2$, and $2\leq N\leq21$. The black solid line represents the equality in Eq.~\eqref{eq:ctur}. For all values of $N$, there are parameter values for which the bound in Eq.~\eqref{eq:ctur} is close to saturated. Parameter values that violate the naive TUR, for which the correlation inference strategy can be applied, correspond to  points that fall to the left of the vertical line in Fig.~\ref{fig:figure3}(a).

\textit{Continuous coupled processes.---}
We next consider several examples that involve $N$ continuous phases $\phi_\alpha$ with $\alpha=1,...,N$, described by systems of coupled Langevin dynamics in the overdamped regime, with the general form
\begin{equation}
    \dot {\phi}_\alpha = \sum_{\beta = 1}^N \left\{ M_{\alpha \beta} \left(  - \partial_\beta \mathcal{U} \right) + \sqrt{2  k_B T} \Sigma_{\alpha \beta} \xi_\beta  \right\},
    \label{eq:langevin}
\end{equation}
with $\mathcal{U}$ a generic potential, $M_{\alpha \beta}$ a mobility matrix with constant coefficients (i.e.~independent of $\phi_\alpha$), $\Sigma_{\alpha \beta}$ the square root of the mobility matrix satisfying $\Sigma_{\alpha \nu} \Sigma_{\beta \nu} = M_{\alpha \beta}$, and $\xi_\beta$ a white noise of unit strength. For the mobility matrix, we set all diagonal coefficients to $M_{\alpha\alpha}=M$ and all off-diagonal coefficients to $M_{\alpha\beta}= h/(N-1)$ ($\alpha \neq \beta$), so that $\tilde{h} \equiv h/M$ is a dimensionless measure of the strength of coupling mediated by the mobility matrix.

We first consider a minimal model of $N=2$ coupled phases that can be treated analytically. This model could describe two driven stochastic gears or rotors, as  represented by the entrained gears in Fig.~\ref{fig:figure_one}(c), in which case the phases $\phi_{1}$ and $\phi_2$  represent the internal state (angular position) of these rotors. The potential $\mathcal{U}$, is chosen as $\mathcal{U}(\phi_1,\phi_2)= -F(\phi_1+\phi_2)  - K\cos(\phi_1-\phi_2) - v\cos(\phi_1+\phi_2)$ where $F$, $K$, and $v$ are arbitrary constants. The first term is the nonequilibrium drive, with a driving force $F$ which is related to the energy dissipation per cycle (when a phase has advanced by $2\pi$), given by $\Delta \mu = 2\pi F$. The second term is a Kuramoto-type coupling that favors synchronization of the two phases \cite{kuramoto1984chemical}. Finally, the third term an anti-synchronizing coupling that favors opposite rotation of the phases and creates energy barriers for the synchronized advances of the two phases. The problem can be solved analytically by a change of variables to the average phase $\Theta=(\phi_1+\phi_2)/2$ and phase difference $\Delta = \phi_1 - \phi_2$ (see Appendix B).
Analytically calculated results for $0<F/k_B T<1$, $0<v/k_B T<2$, $0<K/k_B T<4$, and $\tilde{h}=0.3$ are shown  in Fig.~\ref{fig:figure3}(b). 

We next consider $N$ thermally-activated oscillators that are coupled purely dissipatively, i.e.~only through the off-diagonal components of the mobility matrix, with dimensionless strength $\tilde{h}$. The potential $\mathcal{U}$  is set  to $\mathcal{U}(\{\phi_\alpha\})=  \sum_{\alpha=1}^N V(\phi_\alpha)$ with $V(\phi)=-F \phi  - v\cos(\phi)$ a washboard potential. This model has been shown to provide a description of the dynamics of mechanically coupled enzymes, which become effectively deterministic and synchronized at sufficiently high $\tilde{h}$ \cite{agudo2021synrhonization,Chatzittofi2023,chatzittofi2024topologicalphaselockingmolecular}. The results of numerical simulations of this model for $N=10$, $0<\tilde{h}<8$, $0.4<F/v<0.9$ and $k_B T/v = 0.08, 0.15, 0.2$ are shown in Fig.~\ref{fig:figure3}(c).

Finally, we consider the case of $N$ thermally-activated oscillators with Kuramoto-type coupling  \cite{kuramoto1984chemical}, previously studied in Ref.~\citenum{Shinomoto1986}. In this case, we set $\tilde{h}=0$ so that the mobility matrix is diagonal, and we set $\mathcal{U}(\{\phi_\alpha\})=  \sum_{\alpha=1}^N V(\phi_\alpha) - \frac{K}{N} \sum_{\alpha=1}^N \sum_{\beta=\alpha+1}^N \cos(\phi_\alpha-\phi_\beta)$ where $V(\phi)$ is the same washboard potential as above. The results of numerical simulations  of this model for $N=4$, $0.1<F/v<0.9$, $0.01<k_BT / v<2$, and $K/v=1,6,10$ are shown in Fig.~\ref{fig:figure3}(d).

For all three continuous models [Fig.~\ref{fig:figure3}(b--d)], we find that there are regions of parameter space where the naive TUR is violated, i.e.~$\frac{D}{\Omega}\frac{\Delta \mu}{k_B T} < 1$, and the correlation inference strategy can be applied. In all cases, saturation of the bound in Eq.~\eqref{eq:ctur} is facilitated when the noise strength $k_B T$ is large relatively to the energy barriers whose height is controlled by $v$, as in this case the dynamics become analogous to those of a particle under a constant force, which are known to saturate the TUR. However, to ensure that nonzero correlations survive the couplings must remain sufficiently strong relative to thermal fluctuations. A notable exception is the case of dissipatively-coupled oscillators, Fig.~\ref{fig:figure3}(c), which can violate the naive TUR and come close to saturating the bound even at low noise strength. This can be understood as a consequence of the fact that the dissipative coupling induces quasi-deterministic dynamics even in the absence of noise \cite{agudo2021synrhonization,Chatzittofi2023}.

\textit{Discussion.---}
By applying the MTUR to an ensemble of statistically-identical coupled processes with tight-coupling to entropy production, we have derived a bound (Eq.~\eqref{eq:ctur}) that allows for thermodynamic inference of the strength of correlations and the number of interacting processes in the system, even when only one or a small subset of them is experimentally accessible. {In particular, when only a single current is observed, our strategy provides a  lower bound on $C(N-1)$, where $C$ is the strength of correlations and $N$ the number of interacting processes. When two currents are observed, and thus $C$ can be measured experimentally, our strategy provides a lower bound on $N$.} The inference strategy is applicable when a ``naive'' application of the TUR to a single observable (i.e.~assuming that this observable is isolated or uncorrelated to others) shows an apparent violation. 
By studying a number of minimal toy models that we solved analytically and numerically, we showed that the naive TUR is broken (and thus our proposed inference strategy is applicable) in large portions of parameter space.

One possible way of easily and directly testing the proposed inference strategy experimentally would be in controlled experiments using several driven colloids  in an optical ring [Fig.~\ref{fig:figure_one}(c)]. This experimental setup can produce constant driving forces \cite{hydrodynamic2011sokolov,PhysRevE.90.042302} as well as washboard-like potentials \cite{lee2006giant}. When two or more colloids are present in the ring, hydrodynamic interactions between them can lead to correlations \cite{hydrodynamic2011sokolov}. Otherwise, our proposed strategy could be applied to experiments with teams of molecular motors pulling on the same cargo \cite{nettesheim2020macromolecular} or clustered enzymes catalyzing chemical reactions \cite{xie1999single}.

Finally, we note that, although we have focused here on the inference of correlations provided that the energy dissipation per step ($\Delta \mu$) is known, our results also have implications for the experimental inference of $\Delta \mu$ when it is unknown. Indeed, Eq.~\eqref{eq:ctur} shows that, in an interacting system, individual currents behave as if they were driven by an effective energy dissipation per step $\Delta \mu_\mathrm{eff} =  [1+C(N-1)] \Delta \mu $, with $\Delta \mu_\mathrm{eff} > \Delta \mu $ when $C>0$. In the limit of strong correlations ($C=1$), the system behaves as if every individual current was driven by the total energy dissipation in the system, i.e.~$\Delta \mu_\mathrm{eff} = N \Delta \mu$, as has been reported in previous studies \cite{lee2018thermodynamic,costantini2024thermodynamic}. An inference strategy unaware of existing correlations could therefore lead to a severe overestimation of (the lower bound on) the true $\Delta \mu$. Our results thus suggest that one must be very careful to experimentally rule out possible interactions with other processes before applying thermodynamic inference to entropy production, even when one can assume tight coupling (i.e.~a fixed amount of energy dissipation per step) between the observed current and entropy production.

\textit{Acknowledgements.---}
We acknowledge support from the Max Planck School Matter to Life and the MaxSynBio Consortium which are jointly funded by the Federal Ministry of Education and Research (BMBF) of Germany and the Max Planck Society.

\textit{Appendix A: Analytical solution of discrete model---}
To construct the TUR for this coupled model, we consider the number of steps $\phi_i$ of the $i$-th process. This number can be split into the simultaneous steps $\phi_\mathrm{d}$ that have occurred for all processes due to the diagonal transitions,  and the individual steps $\phi_{\mathrm{s},i}$ taken by each process independently, so that $\phi_i = \phi_\mathrm{d} + \phi_{\mathrm{s},i}$.
Importantly, $\phi_d$ and all the different $\phi_{\mathrm{s},i}$ are governed by one-dimensional biased random walks that are statistically independent of each other.
Using standard results for the biased random walk, we can write $\langle \phi_{\mathrm{s},i} \rangle = (k_+-k_-)t$, $\langle \phi_{\mathrm{d}} \rangle = (h_+-h_-)t$, $\langle \phi_{\mathrm{s},i}^2 \rangle-\langle \phi_{\mathrm{s},i} \rangle^2 = (k_++k_-)t$, and $\langle \phi_{\mathrm{d}}^2 \rangle-\langle \phi_{\mathrm{d}} \rangle^2 = (h_++h_-)t$.
Using the definitions of $\Omega$, $D$, and $C$ given in the main text, and exploiting the statistical independence of $\phi_d$ and all the different $\phi_{\mathrm{s},i}$, we can straightforwardly obtain $\Omega = k_+ + h_+ - k_- - h_-$, $D=(k_+ + h_+ + k_- + h_-)/2$, and $C=(h_+ + h_-)/(k_+ + k_- + h_+ + h_-)$.
Together with the detailed balance conditions, these expressions are used to obtain Eqs.~\eqref{eq:correlation} and \eqref{eq:dphiom} in the main text. 

\textit{Appendix B: Analytical solution of continuous model---}
The Langevin equations in Eq.~\eqref{eq:langevin} are equivalent to the Fokker-Planck equation
\begin{align}
    \partial_t P = \partial_\alpha \bigg[ M_{\alpha \beta} \Big((\partial_\beta \mathcal{U})P + k_B T \partial_\beta P\Big) \bigg],
\end{align}
for the probability $P(\{\phi_\alpha\};t)$, where Einstein summation has been used. For the analytically-solvable model, we have $N=2$ phases, $\mathcal{U}(\phi_1,\phi_2)= -F(\phi_1+\phi_2)  - K\cos(\phi_1-\phi_2) - v\cos(\phi_1+\phi_2)$, $M_{11}=M_{22}=M$ and $M_{12}=M_{21}=h$ as described in the main text.

By performing a linear transformation we change variables to go the average phase $\Theta=(\phi_1+\phi_2)/2$ and phase difference $\Delta = \phi_1 - \phi_2$. The Fokker-Planck equation becomes
\begin{align}
    \partial_t P &=\partial_\theta \bigg[\frac{M+h}{2}\bigg( (\partial_\theta \mathcal{U}) P + k_B T \partial_\theta P \bigg) \bigg] \\&+ \partial_\Delta \bigg[2(M-h)\bigg( (\partial_\Delta \mathcal{U}) P + k_B T \partial_\Delta P \bigg) \bigg], \nonumber
\end{align}
where we have $\mathcal{U}(\Theta,\Delta) = -2F \Theta - v\cos2\Theta - K\cos\Delta$. Because the potential becomes separable in these coordinates, we can obtain separate Fokker-Planck equations for the marginal distributions
\begin{align}
    P_\Theta = \int d \Delta P(\Theta,\Delta),\ \ \ P_\Delta = \int d \Theta P(\Theta,\Delta),
\end{align}
given by
\begin{align}
    \partial_t P_\Theta &= \partial_\Theta \bigg[\frac{M+h}{2} \bigg( [\partial_\Theta V_\Theta(\Theta)] P_\Theta + k_B T \partial_\Theta P_\Theta \bigg) \bigg], \label{eq:Theta} \\
    \partial_t P_\Delta &= \partial_\Delta \bigg[2(M-h) \bigg([\partial_\Delta V_\Delta(\Delta)] P_\Delta + k_B T \partial_\Delta P_\Delta \bigg) \bigg], \label{eq:Delta}
\end{align}
where $V_\Theta(\Theta)=-2F \Theta -v\cos2\Theta$ and $V_\Delta(\Delta)=- K\cos\Delta$.

Equations \eqref{eq:Theta} and \eqref{eq:Delta} each represent the stochastic dynamics of a (driven) particle in a one-dimensional periodic potential. The average velocity and long-time effective diffusion coefficient of a particle in such systems can be calculated analytically, with closed form expressions given in Refs.~\cite{lifson1962self,reimann2001giant,PhysRevE.65.031104} which we do not reproduce here. In the case of Eq.~\eqref{eq:Theta}, the particle is driven by a force $2F$ and one obtains an average velocity $\langle \dot{\Theta} \rangle \neq 0$ and an effective diffusion coefficient $D_\Theta$. In the case of Eq.~\eqref{eq:Delta}, the particle is not driven and thus the average velocity vanishes, $\langle \dot{\Delta} \rangle = 0$, while the effective diffusion coefficient is denoted by $D_\Delta$.

As a final step, we note that $\phi_1$ and $\phi_2$ are related to $\Theta$ and $\Delta$ by the inverse transformations $\phi_1 = \Theta + \Delta/2$ and $\phi_2 = \Theta - \Delta/2$. Exploiting the fact that the dynamics of $\Theta$ and $\Delta$ are statistically independent, we can use the definitions of $\Omega$, $D$, and $C$ given in the main text to obtain $\Omega = \langle \dot{\Theta} \rangle$, $D = D_\Theta + D_\Delta/4$ and $C=(D_\Theta-D_\Delta/4)/(D_\Theta+D_\Delta/4)$.


\providecommand{\noopsort}[1]{}\providecommand{\singleletter}[1]{#1}

\end{document}